\documentclass[
showkeys,
 amsmath,amssymb,
 aps,
Prb,
]{revtex4-2}

\usepackage{graphicx}
\usepackage{dcolumn}
\usepackage{bm}
\usepackage{color}
\usepackage{subfigure}

\begin{document}

\preprint{APS/123-QED}

\title{Non-Local Elastic Lattices with $\mathcal{PT}$-Symmetry and Time Modulation: From Perfect Trapping to the Wave Boomerang Effect}

\author{Emanuele Riva}
\email[]{emanuele.riva@polimi.it}
\affiliation{Department of Mechanical Engineering, Politecnico di Milano, 20156 (Italy)}

\date{\today}

\begin{abstract}
Wave motion is fundamentally constrained by the dispersion properties of the medium, often making it challenging—or even impossible—to guide wave packets along desired trajectories, particularly when wave inversion is required.
The paper illustrates how one-dimensional (1D) and two-dimensional (2D) non-Hermitian elastic lattices with time-varying non-local feedback interactions offer unprecedented wave guidance. 
By relaxing the constraint of Hermiticity while preserving $\mathcal{PT}$-symmetry of the nonlocal interactions, it is herein built a framework where the dispersion transitions from positive to negative group velocity, passing through an intermediate regime characterized by a perfectly flat band across all momenta. This effect, realized within the unbroken $\mathcal{PT}$-symmetric phase, is further enhanced by the time modulation of lattice parameters, thereby unlocking functionalities such as perfect trapping, where a wave packet is intentionally stopped, and the wave boomerang effect, where the wave packet is reversed or guided back to its initial position. The framework presented in this paper unlocks opportunities that extend beyond wave guidance, including information processing through dispersion engineering in elastic media.


\end{abstract}

\keywords{Flat Bands, $\mathcal{PT}$-Symmetry, Non-Hermitian Mechanics, Time-Varying metamaterials, Temporal Waveguiding.  }

\maketitle

\section{Introduction}
A foundational concept in wave physics is the Hermiticity of the Hamiltonian, a mathematical symmetry that guarantees real eigenvalues and ensures energy conservation within the system. 
Yet, the exploration of non-Hermitian systems, which extend beyond this conventional framework, has uncovered a realm of phenomena characterized by complex eigenvalue spectra and non-unitary state evolution \cite{srivastava2015causality}. While non-Hermitian systems may exhibit instability and energy nonconservation, they also enable the realization of dynamics and functionalities that remain inaccessible in purely Hermitian settings, including asymmetric wave propagation and nonreciprocal energy transport. These distinctive features, often realized through engineered gain and loss, have spurred research on non-Hermitian physics across various domains, from quantum mechanics to wave dynamics in elastic and acoustic media \cite{christensen2024perspective}.  Moreover, under specific symmetries, a fine-tuning of gain and loss interactions can give rise to phases with purely real spectra, known as the unbroken $\mathcal{PT}$-symmetric phase, or phases with complex eigenstates, referred to as broken $\mathcal{PT}$-symmetric phase \cite{bender1998real,bender2019pt}. These two regimes are separated by spectral singularities—exceptional points (EPs)—where two or more eigenvalues coalesce \cite{longhi2018parity,longhi2010spectral}. In other words, the interplay between non-Hermiticity and $\mathcal{PT}$-symmetry replaces the conventional requirement of Hermiticity with $\mathcal{PT}$-symmetric invariance, unlocking a range of novel functionalities and phenomena. Within this framework, recent advances in elastic and acoustic metamaterials have transcended traditional limitations imposed by Hermitian Hamiltonians, and experimental realizations of non-Hermitian systems have demonstrated their feasibility with platforms incorporating gain and loss \cite{thomes2024experimental,wu2019asymmetric}. These implementations range from engineered lattices to active metamaterials that leverage feedback control and digital modulation to break conventional symmetry constraints in acoustics \cite{zhu2014pt,shi2016accessing,thevamaran2019asymmetric} and elasticity \cite{hou2018tunable,merkel2018dynamic,hou2018pt}. Notable applications include invisibility \cite{fleury2015invisible}, directional wave propagation \cite{riva2022harnessing,wu2019asymmetric,wu2023engineering}, enhanced sensitivity \cite{rosa2021exceptional,thomes2024experimental}, and topological pumping \cite{riva2021adiabatic}, to name a few.

A promising approach to achieving non-Hermiticity in mechanical systems relies on the implementation of nonlocal feedback interactions. Nonlocality, where elastic connections extend beyond nearest neighbors, provides a powerful framework to manipulate the dispersion properties of such systems. By carefully designing these interactions, a variety of nonreciprocal effects can be engineered, such as the emergence of skin modes \cite{braghini2021non} and asymmetric wave propagation \cite{rosa2020dynamics}. Beyond these nonreciprocal phenomena, nonlocal interactions also give rise to intriguing dispersion features, such as roton and maxon dispersions—regions of local minima and maxima in the dispersion relation, respectively \cite{kazemi2023drawing,wang2022nonlocal,iglesias2021experimental}. These features have been extensively explored in both one-dimensional and two-dimensional lattices, with particular attention given to systems with partially flat dispersion bands \cite{paul2024complete}.

Driven by the fascinating characteristics of $\mathcal{PT}$-symmetric systems and the tunable dispersion features enabled by nonlocal interactions, this study brings these two concepts together. The wave manipulation protocol employed herein relies on 1D and 2D elastic lattices, where the nonlocal interactions are intentionally introduced in the form of gain and loss and implemented as a pair of feedback forces applied to consecutive sites with equal magnitude but opposite signs. This careful balance between gain and loss ensures that the lattice topology maintains a real spectrum while exhibiting distinctly non-Hermitian dynamics characteristic of the unbroken $\mathcal{PT}$-symmetric phase. By tuning the feedback parameters, the dispersion undergoes a controlled transition from positive to negative velocity modes, passing through a remarkable condition where the dispersion becomes entirely flat. Flat dispersion bands, in turn, enable the emergence of intriguing wave phenomena, such as compact localized modes (CLS), which are wave modes spatially confined to an extreme degree \cite{emanuele2024creating,riva2024enhanced,samak2024direct,karki2023non}. However, in the presence of nonlocal interactions, the formation of CLS is inherently restricted, as these interactions induce energy transport outside the localization domain. 
Despite this restriction, the flatness of the bands facilitates perfect energy trapping, wave guidance, or even wave reversal when the lattice parameters are modulated over time. Perfect energy trapping occurs when a wave packet is halted, which requires the energy to lie on a flat dispersion. Achieving this phenomenon is unfeasible in systems based on spatial modulation and local resonance, due to the inherently small but non-zero velocity of resonant modes and the frequency-preserving wavenumber transformations inherent to spatial modulations \cite{santini2022harnessing,riva2024adiabatic,colombi2016seismic,de2020graded,de2020experimental}. In contrast, temporal modulations enable dynamic variation of the frequency content within the waveguide during wave motion, creating conditions where the frequency becomes flat across all momenta or, depending on the modulation parameters, where the wave motion is reversed. Moreover, to guarantee the absence of modulation-induced scattering, the adiabatic theorem is employed to delineate the transition between non-adiabatic and adiabatic configurations, which is necessary for a smooth guidance. This study explores these dynamic features in both 1D and 2D settings, addressing the limitations of alternative modulation strategies for wave guidance and trapping, such as temporal pumps \cite{xia2021experimental,grinberg2020robust}, time-modulated Hermitian systems \cite{santini2023elastic,riva2023adiabatic,SANTINI2024118632}, and spatially modulated media \cite{riva2020edge,dorn2023inverse,dorn2022ray}.

The article is structured as follows: section 2 reports on the dynamics of 1D lattices, including dispersion analyses, limiting conditions for adiabaticity, and numerical simulations where an incident wave is stopped and time-reversed. Section 3 extends this framework to make it applicable to the dynamics of 2D lattices. Finally, Section 4 provides concluding remarks and perspectives.

\section{1D lattices: perfect trapping and the wave boomerang effect}
\begin{figure}[!t]
    \centering
    {\includegraphics[width=1.0\textwidth]{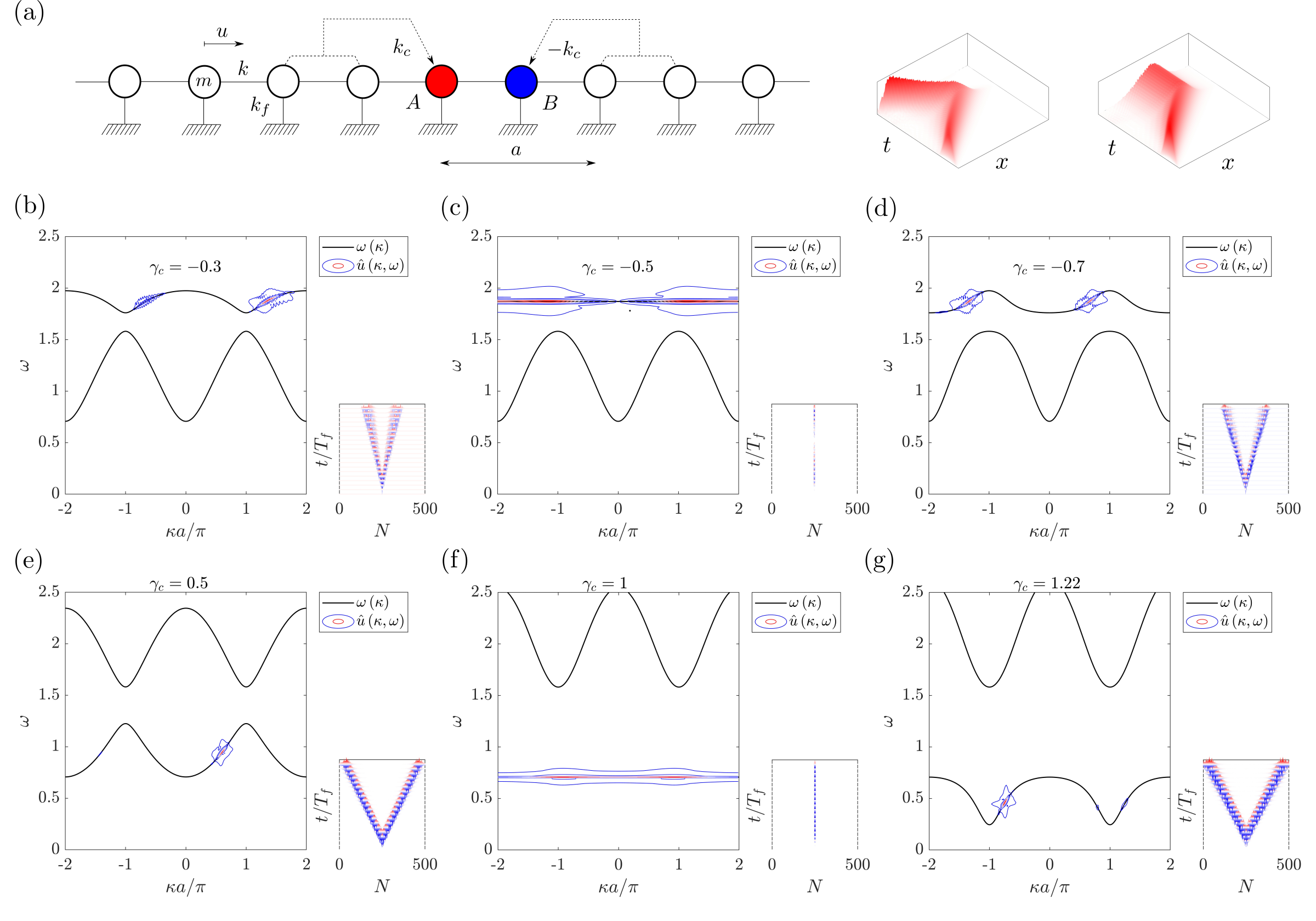}}
    \centering
    \caption{(a) Schematic of the lattice and graphical representation of the temporal trapping and the wave boomerang effect. (b)-(d) Dispersion relations of the lattice under negative $\gamma_c$ values. The black curves are the expected (analytical) dispersion relations, while the contours are computed by Fourier transforming the displacement field measured from numerical simulation. The time history is shown alongside the dispersion plot. The control parameter $\gamma_c$ is set to induce a (b) positive, (c) null, and (d) negative group velocity for a wavenumber impinging on the high-frequency dispersion curve (e)-(g) Dispersion relations of the lattice under positive $\gamma_c$ values, where the control parameter is employed to manipulate the low-frequency dispersion curve from having positive to a negative group velocity. }
    \label{fig:01}
\end{figure}

The discussion begins with the dynamic analysis of the lattice illustrated in Fig. \ref{fig:01}(a), which consists of a set of masses $m=1$ connected by linear time-invariant springs $k=1$ and $k_f=0.5$. These springs link the masses to each other and to the ground, respectively. Beyond these passive interactions, the lattice incorporates active, nonlocal feedback forces with opposite signs $f_{A}=k_c\big[u_B^{p-1}-u_A^{p-1}\big]$ and $f_{B}=-k_c\big[u_B^{p+1}-u_A^{p+1}\big]$, that are applied to the lattice elements A and B of the unit cell. This feedback scheme, which embeds a balanced control gain $k_c$, is systematically patterned across the entire lattice and, as such, the dispersion analysis through Floquet-Bloch boundary conditions is encouraged, starting from the elastodynamic equation of the unit cell:
\begin{equation}
    \begin{split}
        m\ddot{u}_A^{p}+\big(2k+k_f\big)u_A^{p}-k\big(u_B^{p}+u_B^{p-1}\big)-k_c\big(u_B^{p-1}-u_A^{p-1}\big)=0\\[5pt]
        m\ddot{u}_B^{p}+\big(2k+k_f\big)u_B^{p}-k\big(u_A^{p}+u_A^{p+1}\big)-k_c\big(u_A^{p+1}-u_B^{p-1}\big)=0\\        
    \end{split}
    \label{Eq:01}
\end{equation}
where the displacements across consecutive units are linked with complex exponential functions, i.e., $u_{A,B}^{p\pm1}=u_{A,B}^{p}{\rm e}^{\pm{\rm i}\kappa a}$, being $\kappa$ the wavenumber and $a$ the lattice size. 
After performing some algebraic manipulations and assuming harmonic motion, Eq. \ref{Eq:01} can be rewritten in matrix form:
\begin{equation}
    D\big(\omega,\kappa\big)\mathbf{u}=0
\end{equation}
where the dynamical matrix $D(\kappa,\omega)$ is dependent upon frequency $\omega$, the wavenumber $\mu$, the mass, and the stiffness parameters of the lattice:
\begin{equation}
    D\left(\omega,\kappa\right)=\begin{bmatrix}
        \omega_0^2\Big(2+\gamma_f+\gamma_c{\rm e}^{{\rm i}\kappa a}\Big)-\omega^2 &-\omega_0^2\Big(1+{\rm e}^{{\rm i}\kappa a}+\gamma_c{\rm e}^{{\rm i}\kappa a}\Big)\\[8pt]
        -\omega_0^2\Big(1+{\rm e}^{-{\rm i}\kappa a}+\gamma_c{\rm e}^{-{\rm i}\kappa a}\Big)&\;\;\;\omega_0^2\Big(2+\gamma_f+\gamma_c{\rm e}^{-{\rm i}\kappa a}\Big)-\omega^2 
    \end{bmatrix}\hspace{1cm}\mathbf{u}=\begin{pmatrix}        
    u_A^{(p)}\\[8pt]
    u_B^{(p)}
    \end{pmatrix}
    \label{Eq:02}
\end{equation}
here, the eigenvector $\mathbf{u}$ accommodates the motion of the lattice elements A and B, and the key parameters are defined as $\omega_0=\sqrt{k/m}$, $\gamma_f=k_f/k$, and $\gamma_c=k_c/k$. Notably, the eigenvalue problem $D\big(\omega,\kappa\big)\mathbf{u}=0$ leads to the lattice dispersion relation $\omega\big(\kappa\big)$, with the dynamic matrix $D(\kappa,\omega)$ being inherently non-Hermitian as $D^\dagger\neq D$. Due to non-Hermiticity, the time evolution of the wave modes may exhibit non-unitary and potentially unstable behavior. However, the system remains invariant under spatial reflection $x\rightarrow-x$ and time reversal $t\rightarrow-t$ (${\rm i}\rightarrow -{\rm i}$), yielding a phase transition from an unbroken to a broken $\mathcal{PT}$-symmetric phase. While the detailed exploration of this phase transition lies beyond the scope of the present study, this work focuses on the wave phenomena emerging within the unbroken $\mathcal{PT}$-symmetric phase, where the strength of the nonlocal interactions $\gamma_c$ is sufficiently small and, hence, the evolution of the states is bounded. Under these settings, the nonlocal feedback interactions are reminiscent of a perfectly balanced gain and loss, which ensures a purely real spectrum and stable dynamics. Building on this framework, the key idea is to leverage the time modulation of the control parameter $\gamma_c(t)$ to achieve perfect trapping and the boomerang effect of elastic waves, as illustrated schematically in Fig. \ref{fig:01}(a). Perfect trapping occurs when a wave packet is guided toward a desired position while reaching zero group velocity. In the wave boomerang effect, instead, the wave motion is further manipulated to reverse the motion, thereby guiding the wave packet back.

Before addressing the dynamic modulation, the analysis begins with the wave propagation properties of the lattice for fixed $\gamma_c$ values. To this end, the dispersion relation $\omega\big(\kappa\big)$ is evaluated for six distinct and representative conditions, later validated through numerical simulations. Negative values of $\gamma_c$ are reported in Figs. \ref{fig:01}(b)-(d), revealing how the optical (i.e. the high-frequency) dispersion curve transitions from exhibiting a positive group velocity $c_g=\partial\omega/\partial \kappa$, to one with a negative group velocity. Interestingly, for the intermediate case $\gamma_c=-0.5$, the curve becomes completely flat across all momenta, corresponding to a unique condition where the propagating modes have zero group velocity. A similar effect is visible in Figs. \ref{fig:01}(e)-(g) for positive $\gamma_c$ values, this time affecting the acoustic (low-frequency) dispersion curve. 

\begin{figure}[!t]
    \centering
    {\includegraphics[width=0.85\textwidth]{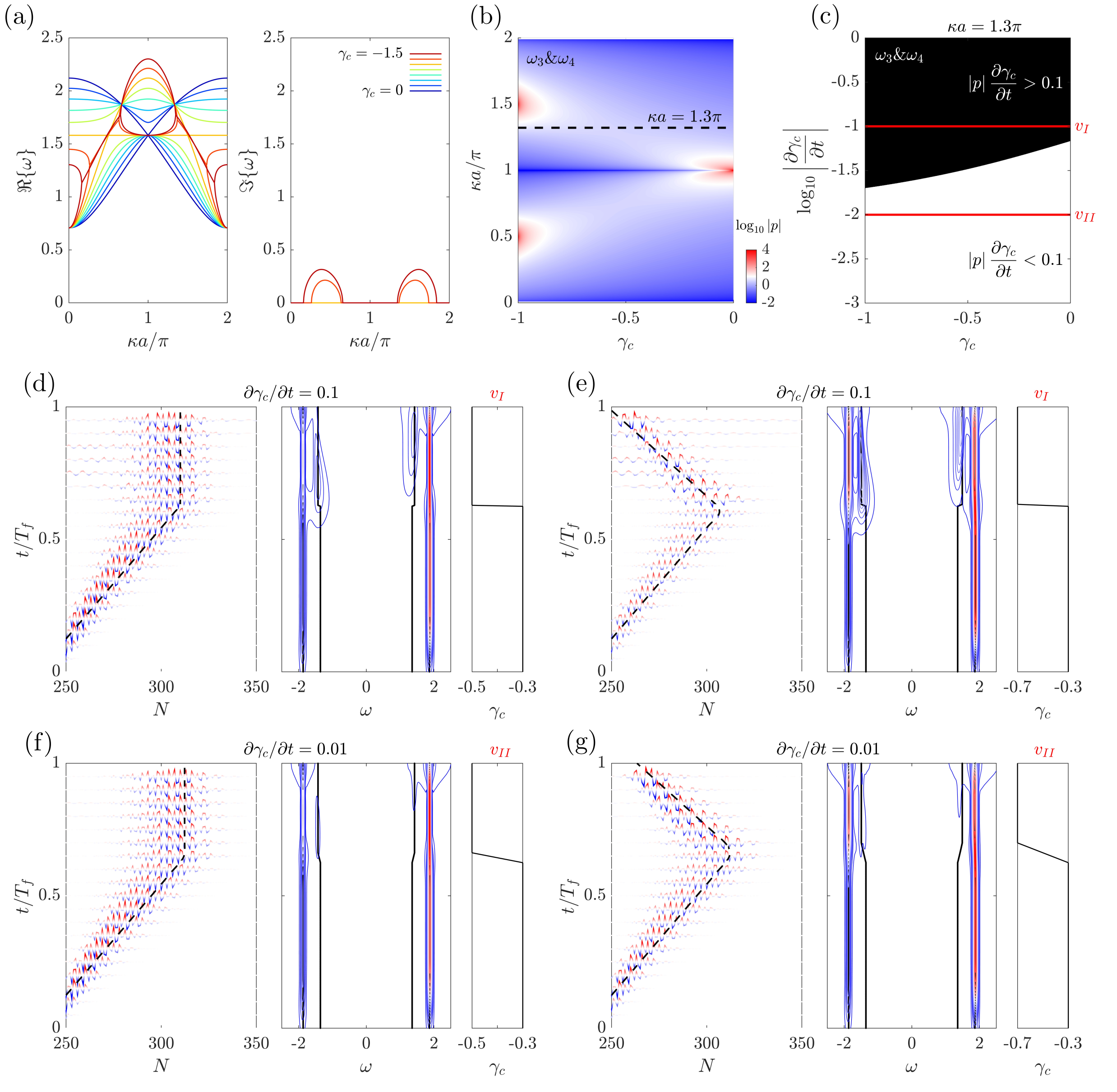}}
    \centering
    \caption{Numerical results upon varying $\gamma_c<0$. (a) Real and imaginary part of the dispersion relation. (b) Coupling term $\log_{10}\big|p_{ij}\big|$ between the wave modes $\omega_3$ and $\omega_4$ for a unitary modulation velocity $\partial\gamma/\partial t$ and for different impinging wavenumbers $\kappa$. (c) Space of parameters $\gamma_c$,$\partial\gamma_c/\partial t$ where the modulation can be considered adiabatic (white region) or nonadiabatic (black region) for an impinging wavenumber $\kappa a=1.3\pi$. The modulation velocities highlighted in red are those employed in the time simulation. 
    (d-e) Time history, corresponding frequency spectrogram, and modulation employed to achieve perfect trapping and the wave boomerang effect. Due to the nonadiabatic nature of the modulation, the energy scattering is visible in figures (d) and (e). Figures (f) and (g) illustrate perfect trapping and the wave boomerang effect under adiabatic conditions. Due to adiabaticity, the scattering is eliminated from both time simulation and frequency spectrogram.    }
    \label{fig:02}
\end{figure}

These features are corroborated through numerical simulations of a lattice made of $N=500$ masses, subjected to a point force excitation at its central site. The input force is a narrowband signal with 20 periods, centered around the flat band frequency. The results are presented as space-time diagrams alongside the corresponding dispersion plots in Figs. \ref{fig:01}(b)-(g). In Figs. \ref{fig:01}(b), \ref{fig:01}(d), \ref{fig:01}(e), and \ref{fig:01}(g), the wave packet propagates toward positive and negative directions, respectively. In contrast, Figs. \ref{fig:01}(c) and \ref{fig:01}(f) illustrate that the wave remains localized near the central position of the lattice and the energy does not spread toward the boundaries of the lattice. While this behavior has similarities to compact localized states (CLS), recently observed in acoustics and mechanics \cite{emanuele2024creating,riva2024enhanced,samak2024direct,karki2023non}, it is fundamentally distinct. CLSs arise from perfect interference effects, producing unitary amplitudes within their domain and vanishing entirely outside it. The system herein considered does not possess extreme and compact localization properties, as a result of the nonlocal interactions that extend from the nearest neighbors. However, the response observed in Figs. \ref{fig:02}(c) and \ref{fig:02}(f) exhibits a remarkable degree of localization, which is due to the group velocity vanishing at flat band frequency. 
To further corroborate these analyses, the time-domain simulations are Fourier transformed, and the resulting numerical dispersions $\hat{u}\big(\kappa,\omega\big)$ are superimposed to the expected theoretical curves, demonstrating a very good agreement. 

The control parameter $\gamma_c<0$ is now varied quasi-statically, producing the set of dispersion relations depicted in Fig. \ref{fig:02}(a). For clarity, the real and imaginary components of the dispersion are shown separately. It can be observed that for $-1<\gamma_c<0$ the spectrum is purely real, while $\gamma_c=-1$ delineates a transition between unbroken and broken $\mathcal{PT}$-symmetric phase. Building on prior studies in the context of temporal waveguiding \cite{santini2023elastic}, $\gamma_c(t)$ is modulated slowly over time, effectively transforming a dispersion curve with positive group velocity into one with either negative or zero slopes. In this context, the adiabatic theorem provides a mathematical procedure to determine whether the energy, injected into a wave mode, can be guided smoothly by following the time evolution of a dispersion curve, achieved through a gradual variation of a relevant parameter, which, in this case, is $\gamma_c$. The first step of this procedure involves recasting Eq. \ref{Eq:01} as a system of first-order differential equations, $\big|\dot{\psi}\big>=H\left(\kappa\right)\left|\psi\right>$, which defines a right eigenvalue problem ${\rm i}\omega\left|\psi^R\right>=H(\kappa)\left|\psi^R\right>$, where $\omega_i(t)$ and $\left|\psi^R_i(t)\right>$ are the time-dependent quasi-static eigenvalues and their corresponding eigenvectors, with $i=1,2,3,4$. 
In the case at hand, for a given wavenumber $\kappa$, the system supports two pairs of counter-propagating wave modes, where each pair has equal magnitude but an opposite sign: $\omega_3=-\omega_2$ and $\omega_4=-\omega_1$. Consistently with the adiabatic theorem, the energy injected into the $i^{\text{th}}$ state remains confined to that state if the following inequality holds: 
\begin{equation}
\left|p_{ij}\frac{\partial\gamma_c}{\partial t}\right|\ll1\hspace{1cm}with\hspace{1cm} p_{ij}=\displaystyle\frac{\left<\psi_i^L\right|\frac{\partial H}{\partial \gamma_c}\left|\psi_j^R\right>}{\left(\omega_i-\omega_j\right)^2}
\label{Eq:03}
\end{equation}
In other words, the $i^{\text{th}}$ wave mode undergoes a transformation governed by the evolution of the $i^{\text{th}}$ dispersion curve, both in terms of its frequency content $\omega_i$ and wave shape $\left|\psi_i^R\right>$, and the term $\big|p_{ij}\frac{\partial\gamma_c}{\partial t}\big|$ quantifies the coupling between mode $i$ and mode $j$. A sufficiently small value ensures that the energy remains confined to $\omega_i$, thereby achieving the modal transformation without energy leak to another wave mode that populates the dispersion for the impinging wavenumber $\kappa$. Mathematical details, other numerical examples, and experiments on this matter are reported in Refs. \cite{santini2023elastic,xia2021experimental,grinberg2020robust,riva2023adiabatic,SANTINI2024118632}.

To prepare numerical simulations with meaningful parameters, the magnitude of $p_{ij}$ is mapped in Fig. \ref{fig:02}(b) as a function of $\gamma_c$ for all incident wavenumbers along the dispersion curve $\omega_4(\kappa)$. Among the available modes, $\omega_4$ is most likely to couple with $\omega_3$, as the latter lies closer in frequency compared to the negative frequency counterparts $\omega_1$ and $\omega_2$. This proximity makes the coupling between $\omega_4$ and $\omega_3$ particularly critical from the perspective of maintaining adiabaticity, especially under two scenarios. First, at $\gamma_c=0$ and $\kappa a=1$, where the optical and acoustic modes intersect; and second, at $\gamma_c=0$ and $\kappa a\approx 0.5$, which corresponds to the unbroken-to-broken phase transition. At this transition point, two distinct eigenvalues coalesce into a non-Hermitian singularity, where the distance $\omega_4-\omega_3$ approaches zero. 
These scenarios are avoided by choosing an incident wavenumber $\kappa a=1.3\pi$ which, for convenience, is highlighted with a horizontal dashed line. For this particular value, the group velocity transitions from positive to negative as $\gamma_c$ varies, and the quantity $\big|p_{ij}\frac{\partial\gamma_c}{\partial t}\big|$ is mapped in Fig. \ref{fig:02}(c) as a function of $\gamma_c$ and $\frac{\partial\gamma_c}{\partial t}$. Without any loss of generality, it is here assumed that the variation of $\gamma_c$ is linear and, hence, $\frac{\partial\gamma_c}{\partial t}$ is constant throughout the temporal modulation. Notably, the white regions in the figure denote the parameter space where the modulation velocity is sufficiently slow to ensure adiabaticity, i.e. when the coupling term in Eq. \ref{Eq:03} remains below a threshold value of $\sim 0.1$. Conversely, the black regions indicate nonadiabatic conditions, where a non-negligible amount of energy leaks from the targeted wave mode to adjacent modes. Two representative modulation velocities $v_I$ and $v_{II}$ are marked in the figure and hereafter employed in the time simulations to assess adiabaticity.

\begin{figure}[!t]
    \centering
    {\includegraphics[width=0.85\textwidth]{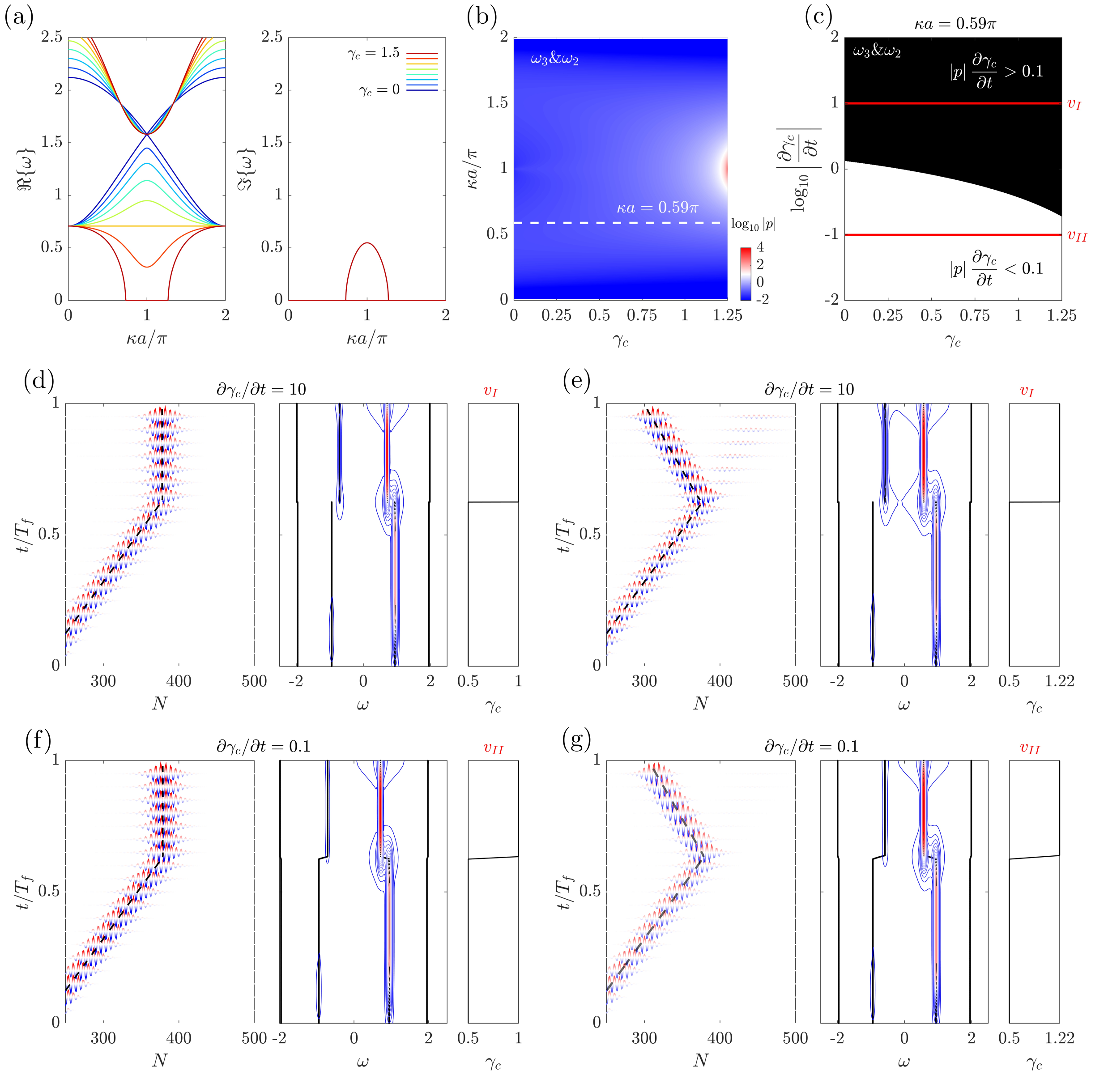}}
    \centering
    \caption{Numerical results upon varying $\gamma_c>0$. (a) Real and imaginary parts of the dispersion relation. (b) Coupling term between $\omega_2$ and $\omega_3$ and (c) corresponding limiting condition for adiabaticity for the impinging wavenumber $\kappa a=0.59\pi$. (d)-(g) Perfect trapping and the wave boomerang effect under (d)-(e) nonadiabatic conditions and (f)-(g) under adiabatic conditions.   }
    \label{fig:03}
\end{figure}
First, a wave packet is generated via a point force applied at the central site of the lattice. The initial nonlocal feedback value is set to $\gamma_c = -0.3$. Time modulation is subsequently employed to either stop the wave packet at a specific lattice position (Figs. \ref{fig:02}(d) and \ref{fig:02}(f)) or reverse its motion (Figs. \ref{fig:02}(e) and \ref{fig:02}(g)). Results are illustrated in terms of the time evolution of the wave packet, a frequency spectrogram, and the applied modulation over time. The frequency spectrogram is computed by windowing the time history of the wave packet, which initially propagates to the right at the frequency $\omega_3$. A moving Gaussian windowing function is employed, $G={\rm e}^{-(t-t_0)^2/c^2}$, whereby the parameter $c$ controls the width of the Gaussian, while $t_0$ varies within the range $[0, T_f]$, being $T_f$ the total simulation duration. The windowed time history is then Fourier transformed to $\hat{u}(\kappa, \omega, t_0)$, and the wavenumber dimension is eliminated by computing the root mean square (RMS) along $\kappa$. The resulting spectrogram $\hat{u}(\omega, t_0)$ is superimposed to the expected evolution of the states $\omega_i$ for the impinging wavenumber $\kappa$. This process captures the time evolution of the frequency content and either validates adiabaticity or reveals energy jumps between distinct wave modes. In addition, the group velocity profile at the impinging wavenumber $\kappa$ is employed to estimate the wave trajectory:
\begin{equation}
    \frac{dx}{dt}=\frac{\partial\omega\left(\kappa,t\right)}{\partial \kappa}
    \label{eq:ray}
\end{equation}
where Eq. \ref{eq:ray} defines a temporal analog of the ray theory developed for spatially varying media \cite{dorn2022ray}. The expected motion of the wave packet, obtained by direct integration of Eq. \ref{eq:ray}, is superimposed to the numerical wave field and represented with a black dashed curve. The results in Fig. \ref{fig:02}(d) illustrates a nonadiabatic condition, where a rapid modulation of $\gamma_c$ from $\gamma_c=-0.3$ to $\gamma_c=-0.5$ drives the wave packet from having a positive to a zero group velocity. As a result, a portion of the energy, primarily injected into $\omega_3$, is halted. While another portion jumps to $\omega_2$ and $\omega_3$, thereby separating the energy content into distinct wave packets. In contrast, Fig. \ref{fig:02}(f) demonstrates a slower modulation velocity, which reduces energy scattering and concentrates the energy more effectively as a zero group velocity mode.
Fig. \ref{fig:02}(e) and Fig. \ref{fig:02}(g) are instead relative to a modulation from $\gamma_c=-0.3$ to $\gamma_c=-0.7$, thereby driving a transition from a positive to a negative group velocity mode. As such, the wave packet, initially propagating from left to right, reverses direction and bends back, effectively inverting its motion. This process occurs in Fig. \ref{fig:02}(e) with significant energy scattering, while Fig. \ref{fig:02}(g) highlights a modulation compliant with the adiabatic theorem, which minimizes energy jumps and ensures a smooth transition. 

The process is repeated, this time targeting the low-frequency dispersion curve, which transitions from positive to negative group velocity modes by varying $\gamma_c$, see Fig. \ref{fig:03}. This transition occurs for values ranging from $\gamma_c = 0$ to $\gamma_c = 1.5$. $\gamma_c = 1$, instead, enables a flat dispersion for all momenta. Under these settings, the adiabaticity is analyzed in Fig. \ref{fig:03}(b) for all impinging wavenumbers and between the wave modes $\omega_3$ and $\omega_2$, which move closer during modulation as compared to $\omega_1$ or $\omega_4$. 
Later, the analysis is focused on the limiting condition for $\kappa a = 0.59\pi$, which is reported in Fig. \ref{fig:03}(c) upon varying $\gamma_c$ and $\partial \gamma_c/\partial t$. The limiting condition is corroborated via time-simulation (Figs. \ref{fig:03}(d-g)) for both nonadiabatic (Figs. \ref{Eq:03}(d-e)) and adiabatic (Figs. \ref{Eq:03}(f-g)) modulations and for both stopped and time-reversed wave packets. Notably, the nonadiabatic case in Fig. \ref{fig:03}(d) does not show any visible modulation-induced wave packet in the time history. However, as shown in the frequency spectrogram, part of the energy jumps toward $\omega_2$ which, for the final value of $\gamma_c=1$, corresponds to a condition where the low-frequency dispersion is flat. Therefore, the scattered wave packet cannot drive away from the impinging wave. Instead, when the target dispersion after modulation has a negative velocity, see Fig. \ref{fig:03}(d), the energy jumps toward the adjacent curve, this time accompanied by a visible wave packet separating from the main ray. Finally, Figs. \ref{fig:03}(f) and \ref{fig:03}(g) illustrate perfect trapping and the wave boomerang effect under adiabatic conditions, where the energy scattering is minimal. 

\section{2D lattices: temporal waveguiding}
\begin{figure}[!t]
    \centering
    {\includegraphics[width=1.0\textwidth]{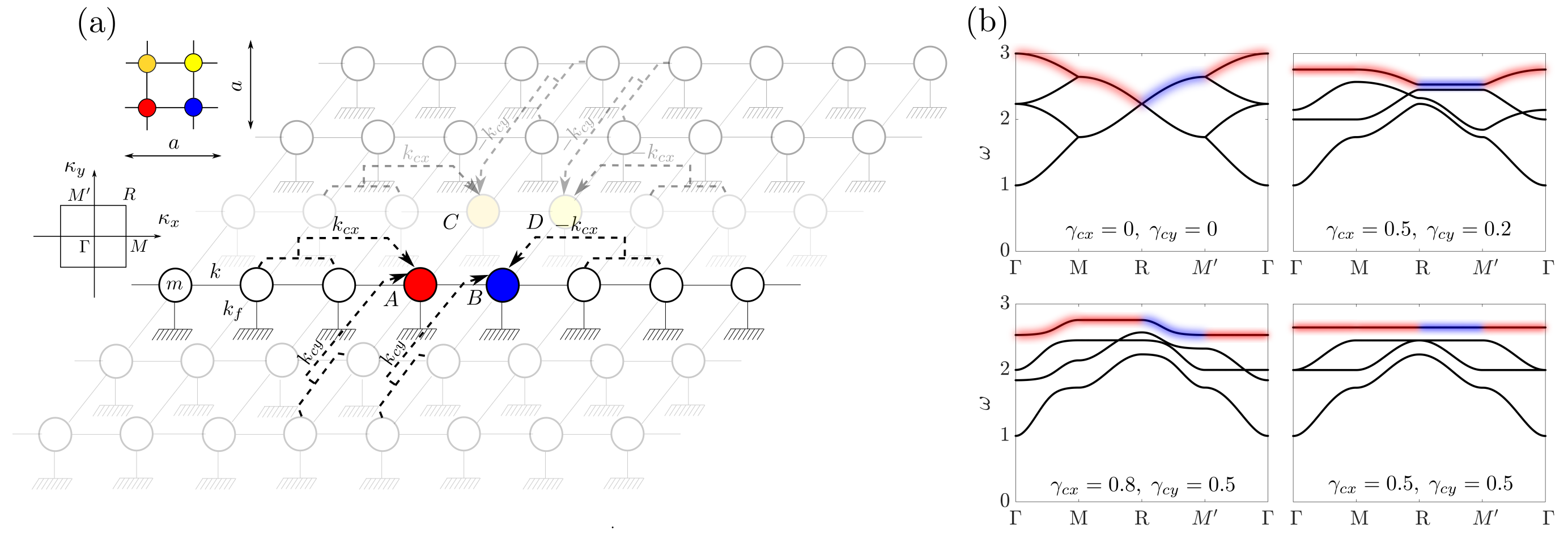}}
    \centering
    \caption{(a) Schematic of the 2D lattice with non-local feedback interactions. The direct and reciprocal unit lattices are shown alongside the schematic. (b) Dispersion curves evaluated along the contours of the Brillouin Zone for different modulation parameters $\gamma_{cx}$, $\gamma_{cy}$. The target dispersion curve is highlighted in red and blue. The blue-highlighted curve illustrates a positive-to-negative transition of the group velocity.}
    \label{fig:04}
\end{figure}

The analysis now focuses on the 2D lattice illustrated in Fig. \ref{fig:04}(a). 
This structure consists of a grid of masses, each connected to the ground and to adjacent masses through springs with spring constants $k_f$ and $k$, respectively. In addition, four nonlocal feedback interactions are applied to the unit cell sites A, B, C, and D:
\begin{equation}
\begin{split}
    f_{A}=k_{cx}\big[u_B^{p-1,q}-u_A^{p-1,q}\big]+k_{cy}\big[u_C^{p-1,q}-u_A^{p-1,q}\big]\hspace{1cm}
    f_{B}=-k_{cx}\big[u_B^{p+1,q}-u_A^{p+1,q}\big]+k_{cy}\big(u_D^{p,q-1}-u_B^{p,q-1}\big]\\[8pt]
    f_{C}=k_{cx}\big[u_D^{p-1,q}-u_C^{p-1,q}\big] - k_{cy}\big[u_C^{p,q+1}-u_A^{p,q+1}\big)\hspace{1cm}f_{D}=-k_{cx}\big[u_D^{p+1,q}-u_C^{p+1,q}\big]-k_{cy}\big[u_D^{p,q+1}-u_B^{p,q+1}\big]
\end{split}    
\end{equation}
Where, in general, the control gains $k_{cx}\neq k_{cy}$.
These feedback forces modify the elastodynamic equations governing the lattice which, in turn, is:
\begin{equation}
\begin{split}
    m\ddot{u}_A^{p,q}+4ku_A^{p,q}-k\left(u_B^{p,q}+u_B^{p-1,q}+u_C^{p,q-1}+u_C^{p,q}\right) + k_{cx}\left(u_B^{p-1,q}-u_A^{p-1,q}\right)+k_{cy}\left(u_C^{p,q-1}-u_A^{p,q-1}\right)=0\\[8pt]    
    m\ddot{u}_B^{p,q}+4ku_B^{p,q}-k\left(u_A^{p,q}+u_A^{p+1,q}+u_D^{p,q}+u_D^{p,q-1}\right) - k_{cx}\left(u_B^{p+1,q}-u_A^{p+1,q}\right)+k_{cy}\left(u_D^{p,q-1}-u_B^{p,q-1}\right)=0\\[8pt]    
    m\ddot{u}_C^{p,q}+4ku_C^{p,q}-k\left(u_A^{p,q}+u_A^{p,q+1}+u_D^{p-1,q}+u_D^{p,q}\right) + k_{cx}\left(u_D^{p-1,q}-u_C^{p-1,q}\right)-k_{cy}\left(u_C^{p,q+1}-u_A^{p,q+1}\right)=0\\[8pt]    
    m\ddot{u}_D^{p,q}+4ku_D^{p,q}-k\left(u_C^{p,q}+u_C^{p+1,q}+u_B^{p,q}+u_B^{p,q+1}\right) - k_{cx}\left(u_D^{p+1,q}-u_C^{p+1,q}\right)-k_{cy}\left(u_D^{p,q+1}-u_B^{p,q+1}\right)=0\\[8pt]        
\end{split}
\label{eq:04}
\end{equation}
with $\gamma_{cx}=k/k_{cx}$ and $\gamma_{cy}=k/k_{cy}$.
From a Bloch analysis perspective, the displacements at neighboring sites are linked through exponential functions $u_{A,B,C,D}^{(p\pm1,q\pm1)}=u_{A,B,C,D}^{(p,q)}{\rm e}^{\pm{\rm i}\big(\kappa_x a+\kappa_y a)}$. 
Under the assumption of harmonic motion and after some algebraic manipulations, Eq. \ref{eq:04} yields an eigenvalue problem $D\left(\omega,\kappa\right)\mathbf{u}=0$, where the dynamic matrix $D(\omega,\kappa)$ is:
\begin{equation}
\begin{split}
    &\hspace{5cm}D\left(\omega,\kappa\right)=\begin{bmatrix}
    D_{11}&D_{12}&D_{13}&0\\[8pt]
    D_{21}&D_{22}&0&D_{23}\\[8pt]
    D_{31}&0&D_{33}&D_{34}\\[8pt]
    0&D_{42}&D_{43}&D_{44}\\ 
    \end{bmatrix} \hspace{1cm} \mathbf{u}=\begin{pmatrix}
        u_A^{p,q}\\[8pt]
        u_B^{p,q}\\[8pt]
        u_C^{p,q}\\[8pt]
        u_D^{p,q}
    \end{pmatrix}\\[8pt]
&D_{11}=\omega_0^2\big(4-\gamma_{cx}{\rm e}^{{\rm i}\kappa_xa}-\gamma_{cy}{\rm e}^{{\rm i}\kappa_ya}\big)-\omega^2 \hspace{0.5cm}D_{12}=\omega_0^2\big(-1-{\rm e}^{{\rm i}\kappa_xa}+\gamma_{cx}{\rm e}^{{\rm i}\kappa_xa}\big)\hspace{0.95cm}D_{13}=\omega_0^2\left(-1-{\rm e}^{{\rm i}\kappa_ya}+\gamma_{cy}{\rm e}^{{\rm i}\kappa_ya}\right)\\[4pt]
&D_{21}=\omega_0^2\big(-1-{\rm e}^{-{\rm i}\kappa_xa}+\gamma_{cx}{\rm e}^{-{\rm i}\kappa_xa}\big)\hspace{0.4cm}D_{22}=\omega_0^2\big(4-\gamma_{cx}{\rm e}^{-{\rm i}\kappa_xa}-\gamma_{cy}{\rm e}^{{\rm i}\kappa_ya}\big)-\omega^2\hspace{0.4cm}D_{24}=\omega_0^2\big(-1-{\rm e}^{{\rm i}\kappa_ya}+\gamma_{cy}{\rm e}^{{\rm i}\kappa_ya}\big)\\[4pt]
&D_{31}=\omega_0^2\big(-1-{\rm e}^{-{\rm i}\kappa_ya}+\gamma_{cy}{\rm e}^{-{\rm i}\kappa_ya}\big)\hspace{0.3cm}D_{33}=\omega_0^2\big(4-\gamma_{cx}{\rm e}^{{\rm i}\kappa_xa}-\gamma_{cy}{\rm e}^{-{\rm i}\kappa_ya}\big)-\omega^2\hspace{0.4cm}D_{34}=\omega_0^2\big(-1-{\rm e}^{-{\rm i}\kappa_xa}+\gamma_{cx}{\rm e}^{{\rm i}\kappa_xa}\big)\\[4pt]
&D_{42}=\omega_0^2\big(-1-{\rm e}^{-{\rm i}\kappa_ya}+\gamma_{cy}{\rm e}^{-{\rm i}\kappa_ya}\big)\hspace{0.2cm} D_{43}=\omega_0^2\big(-1-{\rm e}^{-{\rm i}\kappa_xa}+\gamma_{cx}{\rm e}^{-{\rm i}\kappa_xa}\big)\hspace{0.2cm}D_{44}=\omega_0^2\big(4-\gamma_{cx}{\rm e}^{-{\rm i}\kappa_xa}-\gamma_{cy}{\rm e}^{-{\rm i}\kappa_ya}\big)-\omega^2\\[4pt]
\end{split}
\end{equation}
Now, the dispersion relation $\omega\left(\kappa_x,\kappa_y\right)$ is computed by spanning the contours of the Brillouin zone $\Gamma-M-R-M'-\Gamma$ to qualify the role of the feedback forces on the wave modes supported by the lattice. The results are presented in Fig. \ref{fig:04}(b) for four different sets of parameters, $\gamma_{cx}$ and $\gamma_{cy}$. For convenience, only the positive frequency solutions are displayed, as the dispersion is symmetric about the zero-frequency axis. Notably, for $\gamma_{cx}=0.5$ and $\gamma_{cy}=0.5$, the analysis reveals a high-frequency dispersion band $\omega\left(\kappa_x,\kappa_y\right)$ that is all-flat. This dispersion band, highlighted in blue and red in the figure, is labeled as $\omega_8\left(\kappa_x,\kappa_y\right)$ and hereafter studied more in-depth. If $\gamma_{cx}=0.5$ and $\gamma_{cy}<0.5$, $\omega_8\left(\kappa_x,\kappa_y\right)$ is partially flat, i.e. is flat only along the direction $\Gamma-M$, while $\gamma_{cx}>0.5$ produces a negative group velocity mode along the same direction. This feature will be hereafter employed to stop and control the propagation of wave packets within the lattice. Building upon this, a more intricate modulation law for $\gamma_{cx}$ and $\gamma_{cy}$ is explored. This law extends the modulation amplitude beyond $\gamma_{cx,cy}=0.5$, while ensuring the system remains in the unbroken $\mathcal{PT}$-symmetric phase:
\begin{equation}
\gamma_{cx}=\gamma_{0}\left[1+\alpha\cos{\left(\theta\left(t\right)+\phi\right)}\right]\hspace{1cm}
\gamma_{cy}=\gamma_{0}\left[1+\alpha\sin{\left(\theta\left(t\right)+\phi\right)}\right]
\end{equation}
For simplicity, the time modulation is governed by a linear phase $\theta=\omega_m t$, with characteristic parameters $\gamma_0=0.5$, $\alpha=0.3$, and $\phi=\pi$. 
\begin{figure}[!t]
    \centering
    {\includegraphics[width=1.0\textwidth]{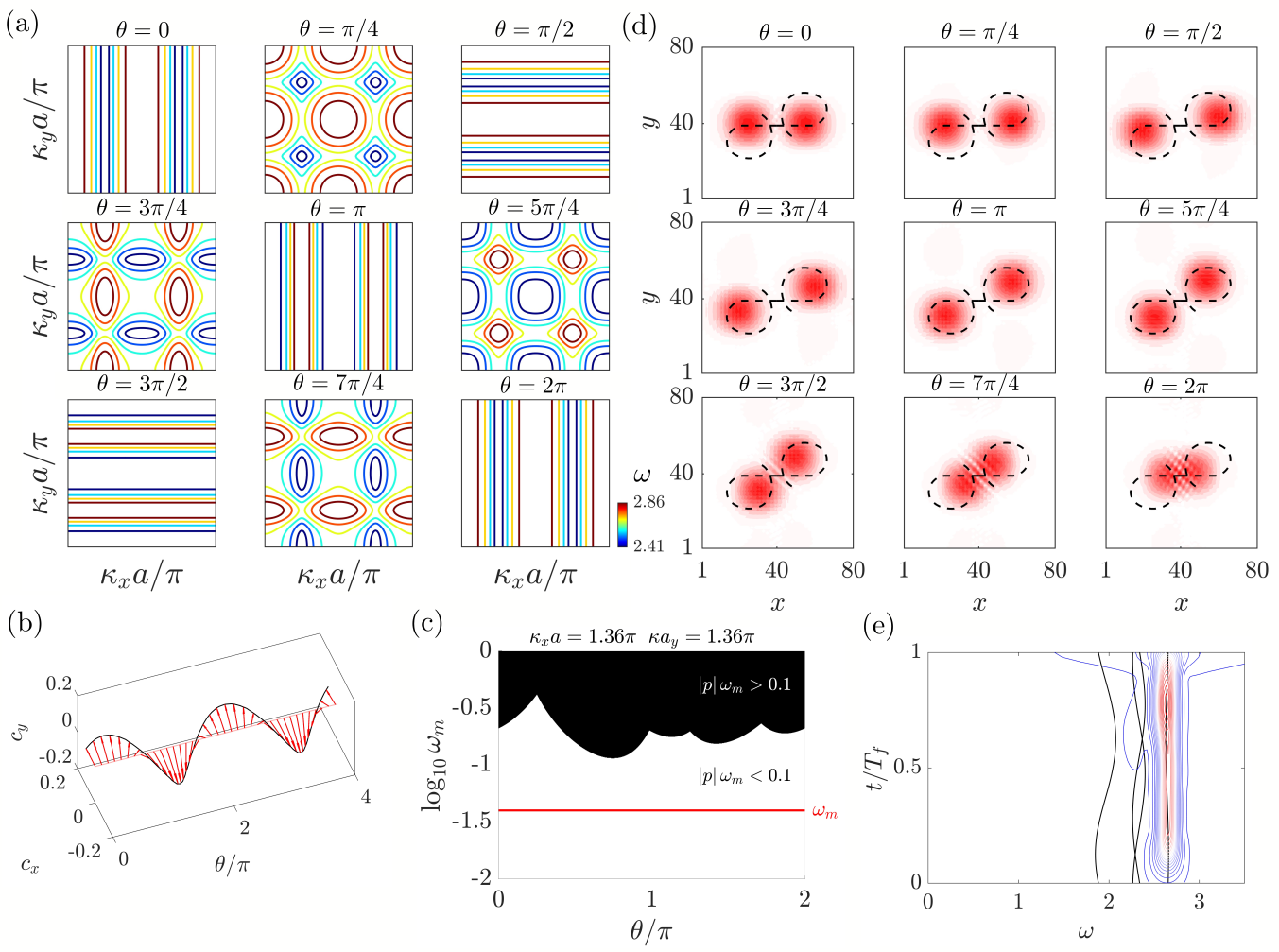}}
    \centering
    \caption{(a) Evolution of the target high-frequency dispersion contours upon varying the parameter $\theta$ and, hence, $\gamma_{cx}$ and $\gamma_{cy}$. (b) Evolution of the group velocity for the impinging wavevector $\mathbf{\kappa}=\left(1.36\pi,1.36\pi\right)$ and (c) corresponding limiting condition of adiabaticity. The adiabatic limit is evaluated between the high-frequency wave mode $\omega_8$ and an arbitrary state $\omega_j$, with $j=1,\ldots,7$. The worst scenario among the adjacent wave modes delimits the black, nonadiabatic, area. The modulation velocity $\omega_m$ employed for simulation is marked with a red horizontal line. (d) Time history for an incident wave packet generated in the middle of the lattice, overlaid with the expected trajectory (dashed line), which is obtained by direct integration of the group velocity. (e) Corresponding spectrogram, demonstrating that the amount of scattering is negligible and, therefore, the transformation can be considered adiabatic.   }
    \label{fig:05}
\end{figure}
Under these modulation settings, the high-frequency dispersion $\omega_8\left(\kappa_x,\kappa_y\right)$, depicted in Fig. \ref{fig:05}(a), undergoes a remarkable transformation by which the isofrequency contours become highly directional, and the directionality is tunable with $\gamma_{cx}$ and $\gamma_{cy}$. Specifically, Fig. \ref{fig:05}(a) illustrates different modulation instants, evaluated for equally spaced $\theta$ values within the range $\left[0,2\pi\right]$, thereby corroborating the control-induced directionality. This cyclical variation is accompanied by changes in the group velocity vector $\mathbf{c}=\nabla \omega$, which is reported in Fig. \ref{fig:05}(b) for an incident wavevector $\mathbf{\kappa}=\left(1.36\pi,1.36\pi\right)$. Without any loss of generality, this target wave mode is hereafter employed for temporal wave guidance. 
Indeed, the group velocity illustrated in Fig. \ref{fig:05}(b) undergoes a dynamic transition, switching from positive to negative and back to positive values. This feature generates a circular trajectory, enabling smooth and continuous redirection of the incident wave packet through a full $360^\circ$ rotation (see the animations provided in the supplementary material). 
Now, to establish an adiabatic transformation where $\theta$ varies over time, the dynamic equation is reformulated into a set of first-order ordinary differential equations, enabling the application of the adiabatic theorem described in Eq. \ref{Eq:03}. 
As highlighted in the dispersion plots, the 2D lattice supports four pairs of counter-propagating waves, and, to leverage the aforementioned directionality, the energy is injected into the high-frequency dispersion curve $\omega_8(\kappa_x,\kappa_y)$. To activate an adiabatic modulation, this wave mode must not couple with neighboring modes, thereby assuming the inequality in Eq. \ref{Eq:03} is satisfied for any wave mode populating the lattice. As such, the most restrictive limiting condition for adiabaticity between mode $\omega_i$ and $\omega_j$ (where $i=8$ and $j=1,2,\ldots,7$) is illustrated in Fig. \ref{fig:05}(c) by taking the maximum value among all $p_{ij}$. Later, a modulation velocity $\omega_m$ is selected for the simulation, ensuring adiabatic guidance of the incident wave packet. 
\begin{figure}[!t]
    \centering
    {\includegraphics[width=1.0\textwidth]{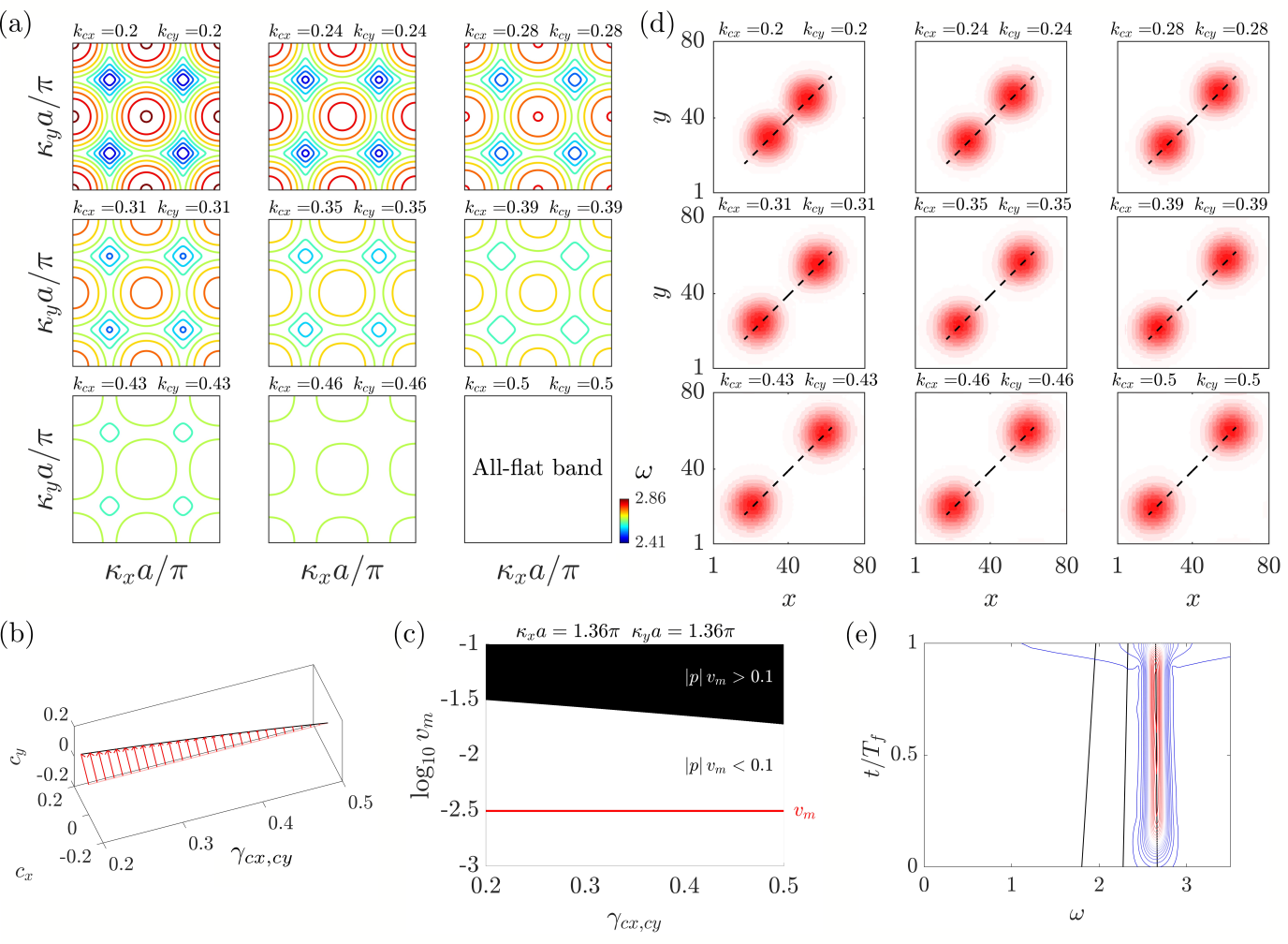}}
    \centering
    \caption{(a) Evolution of the target high-frequency dispersion contours upon varying the parameters $\gamma_{cx}$,$\gamma_{cy}$. (b) Evolution of the group velocity profile for the impinging wavenumbers $\kappa_xa=1.36\pi$ and $\kappa_ya=1.36\pi$ and (c) corresponding limiting condition of adiabaticity. The modulation velocity $v_m$ employed for simulation is marked with a red horizontal line. (d) Time history for an incident wave packet generated in the middle of the lattice, overlaid with the expected trajectory (dashed line), which is obtained by direct integration of the group velocity and (e) corresponding spectrogram. }
    \label{fig:06}
\end{figure}

Simulation-wise, $\theta$ is initially set to zero, allowing the wave packet to propagate along the direction dictated by the initial group velocity (i.e., along $x$). Time modulation is then activated, and the resulting time evolution—also provided as a supplementary animation—is shown in Fig. \ref{fig:05}(d). The figure highlights nine key time instants, overlaid with a dashed curve representing the expected trajectory derived by integrating the group velocity over time. Remarkably, the wave packet closely follows this trajectory, ultimately returning to its initial position through the time modulation, thereby realizing the wave boomerang effect. It can be also observed that the process occurs adiabatically, with minimal scattering. As such, the spectrogram in Fig. \ref{fig:05}(e) confirms the absence of undesired energy transfer from the high-frequency dispersion mode to the neighboring modes, thereby corroborating adiabaticity.

The analysis is repeated by applying a linear modulation from $\gamma_{cx} = \gamma_{cy} = 0.2$ to $\gamma_{cx} = \gamma_{cy} = 0.5$, which facilitates a decrease in velocity to zero, creating a condition where the dispersion $\omega_8(\kappa_x, \kappa_y)$ becomes completely flat:
\begin{equation}
    \gamma_{cx,cy} = \gamma_{c0}+v_m\left(t-t_0\right)
\end{equation}
where $v_m$ represents the modulation velocity. The time evolution of the dispersion surface is shown in Fig. \ref{fig:06}(a), depicting control-induced contours that preserve the directionality of the impinging wavevector while flattening the surface as $\gamma_{cx} = \gamma_{cy} = 0.5$. For $\mathbf{\kappa} = \left(1.36\pi, 1.36\pi\right)$, the group velocity profile in Fig. \ref{fig:06}(b) shows a reduction in modulus to zero, with the angle $\arctan{(c_y/c_x)}$ remaining approximately constant. Also in this finale example, the transformation is considered adiabatic and, therefore, scattering-free when $v_m$ is chosen sufficiently low, i.e., below the threshold presented in Fig. \ref{fig:06}(c). Under these conditions, the time history shown in Fig. \ref{fig:06}(d) illustrates a wave packet initially propagating along a $45^\circ$ angle. Following the time modulation, the direction of the wave packet remains unchanged, but its velocity is nullified, resulting in perfect rainbow trapping without energy transitions between the target mode and adjacent modes, as depicted in Fig. \ref{fig:06}(e).

\section{Conclusions}
This study has demonstrated how non-Hermitian elastic lattices, engineered with time-varying nonlocal feedback interactions, can fundamentally reshape wave propagation. By relaxing the constraint of Hermiticity while preserving $\mathcal{PT}$-symmetry, these systems exhibit unique dispersion properties, including transitions between positive and negative group velocities and a remarkable flat band regime. These features enable functionalities challenging to access in elastic systems, such as perfect energy trapping, wave reversal, and precise trajectory control of wave packets. 
By offering a tunable platform for wave manipulation, this work paves the way for advanced applications in elastic and acoustic waveguides. Potential avenues for future exploration include extending these principles to higher-order interactions, multi-physical couplings, and experimental implementations in complex media. Also, the framework developed herein represents a significant step toward the realization of next-generation technologies for wave-based energy control and information processing.

\nocite{*}

\section*{References}
\bibliography{apssamp}
 
\end{document}